\documentclass[aps,prb,twocolumn,superscriptaddress]{revtex4-1}

\usepackage{graphicx}
\usepackage[caption=false]{subfig}

\usepackage{nopageno, tikz, amsmath, amsfonts, mathrsfs}

\begin{document}

\title{Quantization of inductively-shunted superconducting circuits}

\author{W.~C.~Smith}
\affiliation{Departments of Applied Physics and Physics, Yale University, New Haven, CT 06520, USA}
\author{A.~Kou}
\affiliation{Departments of Applied Physics and Physics, Yale University, New Haven, CT 06520, USA}
\author{U.~Vool}
\affiliation{Departments of Applied Physics and Physics, Yale University, New Haven, CT 06520, USA}
\author{I.~M.~Pop}
\affiliation{Departments of Applied Physics and Physics, Yale University, New Haven, CT 06520, USA}
\affiliation{Physikalisches Institut, Karlsruhe Institute of Technology, Karlsruhe 76131, Germany}
\author{L.~Frunzio}
\affiliation{Departments of Applied Physics and Physics, Yale University, New Haven, CT 06520, USA}
\author{R.~J.~Schoelkopf}
\affiliation{Departments of Applied Physics and Physics, Yale University, New Haven, CT 06520, USA}
\author{M.~H.~Devoret}
\affiliation{Departments of Applied Physics and Physics, Yale University, New Haven, CT 06520, USA}

\date{\today}

\begin{abstract}

We present a method for calculating the energy levels of superconducting circuits that contain highly anharmonic, inductively-shunted modes with arbitrarily strong coupling. Our method starts by calculating the normal modes of the linearized circuit and proceeds with numerical diagonalization in this basis. As an example, we analyze the Hamiltonian of a fluxonium qubit inductively coupled to a readout resonator. While elementary, this simple example is nontrivial because it cannot be efficiently treated by the method known as ``black-box quantization,'' numerical diagonalization in the bare harmonic oscillator basis, or perturbation theory. Calculated spectra are compared to measured spectroscopy data, demonstrating excellent quantitative agreement between theory and experiment.

\end{abstract}


\maketitle

\section{Introduction\label{sec:intro}}

Superconducting qubits provide a promising platform for building a quantum computer, but experimental design requires highly accurate relationships between fabrication parameters and Hamiltonian parameters.\cite{Devoret2013} Successful approaches to this problem have involved finding effective superconducting circuit Hamiltonians with sufficient detail to accurately predict device behavior that are still simple enough for efficient diagonalization.\cite{Yurke1984,Devoret1997,Vion2002,Chiorescu2003,Burkard2004,Blais2004,Chiorescu2004,Wallraff2004,Devoret2004,Koch2007,Manucharyan2009}

The method known as ``black-box quantization,'' in particular, decomposes a distributed microwave environment containing weakly anharmonic superconducting qubits into a few-body effective Hamiltonian.\cite{Nigg2012} It then numerically calculates the normal modes of the system in the limit that the Josephson junctions become linear inductors, writes the Hamiltonian in these normal coordinates, and truncates the Taylor expansions of the cosine terms. This method removes arbitrarily strong coupling terms, but with two caveats: (i) a finite Foster equivalent circuit, used to write a Hamiltonian from the normal modes, is not always accurate and (ii) truncating the cosine expansions is invalid for highly anharmonic qubits (Fig.~\ref{fig:circuit}a), by definition. The first issue has been resolved\footnote{A major difference between Ref.~\onlinecite{Nigg2012} and \onlinecite{Solgun2014} is that the former includes the linear part of the Josephson inductance in the linearized circuit while the latter does not.} using a Brune equivalent circuit.\cite{Solgun2014} In this article, we propose and experimentally confirm a solution to the second issue.

Highly anharmonic qubits are prized for their \emph{in situ} tunability and rich energy level structure, but the numerical diagonalization of their Hamiltonians is costly. A proper choice of basis for numerical diagonalization, which reflects the underlying physics of the system, is paramount to avoid disastrous convergence problems. For example, charge bases capture the band structure of capacitively-shunted qubits like the Cooper pair box, while bound states in inductively-shunted qubits like the fluxonium are conveniently described by harmonic oscillator bases.\cite{Girvin2014} Restricting our attention to the latter class, an early approach was to numerically diagonalize the full Hamiltonian, describing the qubit and the environment, in the bare harmonic oscillator basis.\cite{Manucharyan2009} An alternative method involves numerically diagonalizing the qubit Hamiltonian and treating its coupling to environmental normal modes perturbatively, which has limited applicability when this coupling is strong.\cite{Zhu2013.1} Our strategy is to use the second quantized Fock basis for numerical diagonalization of the full Hamiltonian, after a transformation to the first quantized dressed mode basis. Our central result is that this method is advantageous when the coupling is sufficiently large.

This article demonstrates the application of this strategy to a fluxonium qubit inductively coupled to a readout resonator (Sec.~\ref{sec:device}), and it is organized as follows. In Sec.~\ref{sec:theory}, we linearize an effective electrical circuit of the system, transform into normal coordinates, and numerically diagonalize the Hamiltonian in this basis. Sec.~\ref{sec:experiment} examines the agreement between the calculated energy levels and experimental results. 

\section{Physical device\label{sec:device}}

\begin{figure}
\includegraphics[width=0.48\textwidth]{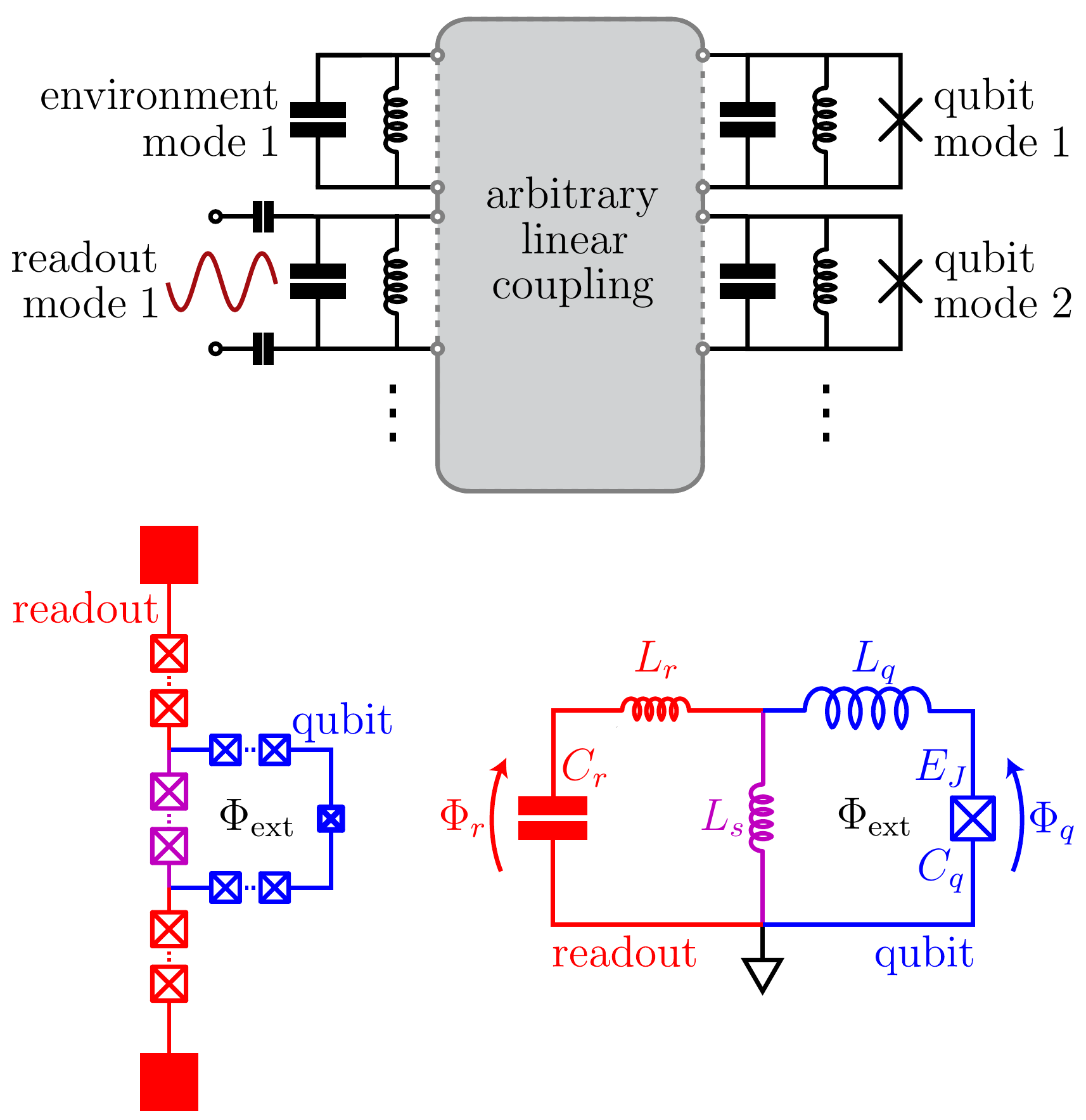}
\put(-244,250){(a)}
\put(-244,130){(b)}
\put(-153,130){(c)}
\caption{(a) Schematic diagram of cQED systems involving a multi-mode microwave environment on the left and various highly anharmonic, inductively-shunted qubits on the right. (b) Schematic diagram of the fluxonium-resonator system. The fluxonium qubit (blue) is threaded by an external magnetic flux $\Phi_\text{ext}$ and coupled to a dipole antenna used as a readout resonator (red) via an array of shared Josephson junctions (purple). (c) Circuit model of the system, obtained by replacing junction arrays with linear inductors. \label{fig:circuit}}
\end{figure}

Experimentally, the fluxonium qubit is realized as one small Josephson junction in parallel with a series array of $\sim$\,100 larger Josephson junctions.\cite{Manucharyan2009} In the limit of large size difference between these two types of junctions, the capacitances and nonlinear inductances across the array junctions may be neglected, and they form a collective, linear superinductance of up to $\sim$\,$300~\text{nH}$.\cite{Masluk2012} If an external magnetic flux is threaded through this loop with magnitude $\Phi_\text{ext} = 0.5\Phi_0$ (with $\Phi_0$ the magnetic flux quantum), the two lowest energy quantum states are approximately the symmetric and antisymmetric superpositions of counter-circulating persistent-current states, similar to the flux qubit.\cite{Chiorescu2003,Manucharyan2009} This feature of the low-energy manifold motivates our use of inductive coupling as the link between the fluxonium qubit and readout resonator. This additional resonator is required for dispersive measurement and, in our system, is an LC oscillator whose capacitance and inductance are formed by an electric dipole antenna and another Josephson junction array, respectively. The inductive coupling is achieved by sharing junctions between the arrays of both oscillators. The resulting fluxonium-resonator system is depicted by the schematic diagram in Fig.~\ref{fig:circuit}b.

To analyze this system, we (i) replace the distributed dipole antenna capacitance by a lumped element capacitor, and (ii) replace the large Josephson junction arrays by linear inductors. Approximation (i) is justified by the antenna length ($\sim$\,1$-$$2~\text{mm}$) being much smaller than the wavelength of the resonator ($\sim$\,3$-$$6~\text{cm}$).\cite{Pop2014} On the other hand, (ii) is permitted when all array junctions are much larger than the small junction of the qubit and the relevant frequencies are below the fundamental mode of the array.\cite{Masluk2012,Ferguson2013} Figure \ref{fig:circuit}c shows the resulting electrical circuit.

\section{Theory\label{sec:theory}}

\subsection{Linearization\label{subsec:linear}}

Having obtained a circuit with five elements, we may use Kirchoff's laws to write the Lagrangian in terms of two degrees of freedom: the flux across the capacitance of the readout resonator, $\Phi_r$, and that across the small junction of the qubit, $\Phi_q$. In the limit that $L_q \gg L_r \sim L_s$, i.e.~the unshared inductance of the qubit is much larger than either the unshared inductance of the readout resonator or the shared inductance, the circuit is well-described by the Lagrangian

\begin{align} \label{eq:lagrangian}
\mathcal{L} &= \frac{1}{2} C_r \dot{\Phi}_r^2 - \frac{1}{2 (L_r + L_s)} \Phi_r^2 + \frac{L_s}{L_q (L_r + L_s)} \Phi_r \Phi_q \nonumber\\
&\hspace{1cm}+ \frac{1}{2} C_q \dot{\Phi}_q^2 - \frac{1}{2 L_q} \Phi_q^2 + E_J \cos \left( \varphi_q - \varphi_\text{ext} \right),
\end{align}

\noindent
upon $\Phi_q \longrightarrow \Phi_q - \Phi_\text{ext}$. Here, $C_q$ and $E_J$ denote the capacitance and Josephson tunneling energy of the small junction, while $C_r$ represents the dipole capacitance of the readout mode. We have also introduced $\varphi_q \equiv 2\pi \Phi_q / \Phi_0$ as the difference in arguments of the superconducting order parameter on either side of the small junction. Analogously, we have defined $\varphi_\text{ext} \equiv 2\pi \Phi_\text{ext} / \Phi_0$. Note that the first two terms in (\ref{eq:lagrangian}) describe the readout mode and the last three terms correspond to the qubit mode. All coupling between these modes is captured by the bilinear mutually inductive third term.

To linearize (\ref{eq:lagrangian}), we must temporarily dispense with components of the cosine term. For weakly anharmonic, capacitively-shunted qubits, the cosine can be replaced by the quadratic term in its Taylor expansion.\cite{Nigg2012} An example is the transmon, which is a Josephson junction shunted by a large capacitance, and whose dynamics are governed by multiple Cooper pair tunneling events.\cite{Koch2007} For highly anharmonic, inductively-shunted qubits, the cosine should be discarded altogether. The fluxonium belongs to this class, owing to its large inductive shunt and single Cooper pair charging effects.

\subsection{Normal modes\label{subsec:normal}}

We decouple the linearized Lagrangian ($E_J \rightarrow 0$ in (\ref{eq:lagrangian})) by diagonalizing the classical equations of motion. This amounts to changing variables to $\Phi_R$ and $\Phi_Q$ such that $\Phi_r = \lambda_1 \Phi_R + \lambda_2 \Phi_Q$ and $\Phi_q = \lambda_3 \Phi_R + \lambda_4 \Phi_Q$. The Lagrangian is now given by

\begin{align} \label{eq:normallagrangian}
\mathcal{L} &= \frac{1}{2} C_R \dot{\Phi}_R^2 - \frac{1}{2 L_R} \Phi_R^2 + \frac{1}{2} C_Q \dot{\Phi}_Q^2 - \frac{1}{2 L_Q}\Phi_Q^2 \nonumber \\
&\hspace{1cm} + E_J \cos \left( \lambda_3 \varphi_R + \lambda_4 \varphi_Q - \varphi_\text{ext} \right),
\end{align}

\noindent
where we have defined $\varphi_i \equiv 2\pi \Phi_i / \Phi_0$ to be the normal mode superconducting phases (with $i = R,Q$). Similarly, $C_i$ and $L_i$ denote the normal mode capacitances and inductances given by

\begin{align} \label{eq:normalparams}
C_R &= \lambda_1^2 C_r + \lambda_3^2 C_q \\
C_Q &= \lambda_2^2 C_r + \lambda_4^2 C_q \\
\frac{1}{L_R} &= \frac{\lambda_1^2}{L_r + L_s} + \frac{\lambda_3^2}{L_q} - \frac{2 \lambda_1 \lambda_3 L_s}{L_q (L_r + L_s)} \\
\frac{1}{L_Q} &= \frac{\lambda_2^2}{L_r + L_s} + \frac{\lambda_4^2}{L_q} - \frac{2 \lambda_2 \lambda_4 L_s}{L_q (L_r + L_s)}
\end{align}

\noindent
We note that this normal mode basis, in which the coupling between modes is entirely captured by the cosine term in (\ref{eq:normallagrangian}), conveniently makes obvious the inherited nonlinearity of the readout as well as the nonlinear nature of the coupling.\cite{Nigg2012} To draw an analogy to the circuit quantum electrodynamics (cQED) literature, we observe that $|\lambda_1| \gg |\lambda_2|$ and $|\lambda_3| \ll |\lambda_4|$ in our devices (see Tab.~\ref{tab:lambda}), which means that mode $R$ is vastly more linear than mode $Q$.\footnote{Strictly speaking, this requires $\sqrt{L_R / C_R} \lesssim \sqrt{L_Q / C_Q}$ so that the magnitudes of $\varphi_R$ and $\varphi_Q$ in (\ref{eq:normallagrangian}) are at most comparable.} This allows us to refer to modes $R$ and $Q$ in (\ref{eq:normallagrangian}) as the \emph{readout-like} and \emph{qubit-like} modes, respectively, which we will henceforth refer to as \emph{readout} and \emph{qubit}.


We then apply a Legendre transform and define the conjugate charges $Q_i \equiv \partial \mathcal{L} / \partial \dot{\Phi}_i$ for the two normal modes.\cite{Devoret1997} These steps permit the writing of the quantum Hamiltonian

\begin{align} \label{eq:circuithamiltonian}
\hat{H} &= \frac{1}{2 C_R} \hat{Q}_R^2 + \frac{1}{2 L_R} \hat{\Phi}_R^2 + \frac{1}{2 C_Q} \hat{Q}_Q^2 + \frac{1}{2 L_Q} \hat{\Phi}_Q^2 \nonumber \\
&\hspace{1cm} - E_J \cos \left( \lambda_3 \hat{\varphi}_R + \lambda_4 \hat{\varphi}_Q - \varphi_\text{ext}\right).
\end{align}

\noindent

\subsection{Numerical diagonalization\label{subsec:diag}}

We express this Hamiltonian in the normal mode harmonic oscillator basis $\left\{ |nm\rangle\right\}$, where $n$ and $m$ correspond to the number of excitations in the readout and qubit modes, which yields the (infinite-dimensional) matrix 

\begin{align} \label{eq:matrixhamiltonian}
\hat{H} &= \hbar \sum\limits_{n, m \in \mathbb{N}} \left( \omega_R n + \omega_Q m \right) |nm\rangle \langle nm | \nonumber \\
&\hspace{0.1cm}- E_J \sum\limits_{n, n^\prime, m, m^\prime \in \mathbb{N}} \Big[ \cos \varphi_\text{ext} \left( c^R_{nn^\prime} c^Q_{mm^\prime} + s^R_{nn^\prime} s^Q_{mm^\prime} \right) \nonumber \\
&\hspace{0.2cm}+ \sin \varphi_\text{ext} \left( c^R_{nn^\prime} s^Q_{mm^\prime} - s^R_{nn^\prime} c^Q_{mm^\prime} \right) \Big] |nm\rangle \langle n^\prime m^\prime |.
\end{align}

\noindent
Here, $\omega_i$ is the dressed harmonic oscillator frequency of normal mode $i$. The cosine and sine matrix elements can be computed analytically as follows, where $\mathscr{L}_a^b$ are the associated Laguerre polynomials and it is understood that $\lambda_R = \lambda_3$ and $\lambda_Q = \lambda_4$.\cite{Gradshteyn2007}

\begin{widetext}
\begin{align} \label{eq:cosine}
c^i_{k\ell} \equiv \langle k \left| \cos (\lambda_i \hat{\varphi}_i) \right| \ell \rangle &= \begin{cases} (-1)^{\frac{\ell-k}{2}}\sqrt{\dfrac{k!}{\ell!}} \left(\lambda_i \varphi_i^\text{ZPF} \right)^{\ell - k}  e^{-(\lambda_i \varphi_i^\text{ZPF})^2/2} \mathscr{L}_k^{\ell - k} \left( (\lambda_i \varphi_i^\text{ZPF})^2\right) \hspace{0.28cm} & (k+\ell\text{ even, } k \leq \ell) \\ 0 & (k + \ell\text{ odd}) \end{cases} \\
\label{eq:sine}
s^i_{k\ell} \equiv \langle k \left| \sin (\lambda_i \hat{\varphi}_i) \right| \ell \rangle &= \begin{cases} 0 & (k+\ell\text{ even}) \\ (-1)^{\frac{\ell-k+1}{2}}\sqrt{\dfrac{k!}{\ell!}} \left(\lambda_i \varphi_i^\text{ZPF} \right)^{\ell - k}  e^{-(\lambda_i \varphi_i^\text{ZPF})^2/2} \mathscr{L}_k^{\ell - k} \left(  (\lambda_i \varphi_i^\text{ZPF})^2\right) & (k+\ell\text{ odd, } k \leq \ell) \end{cases}
\end{align}
\end{widetext}

\begin{figure}
\includegraphics[width=0.49\textwidth]{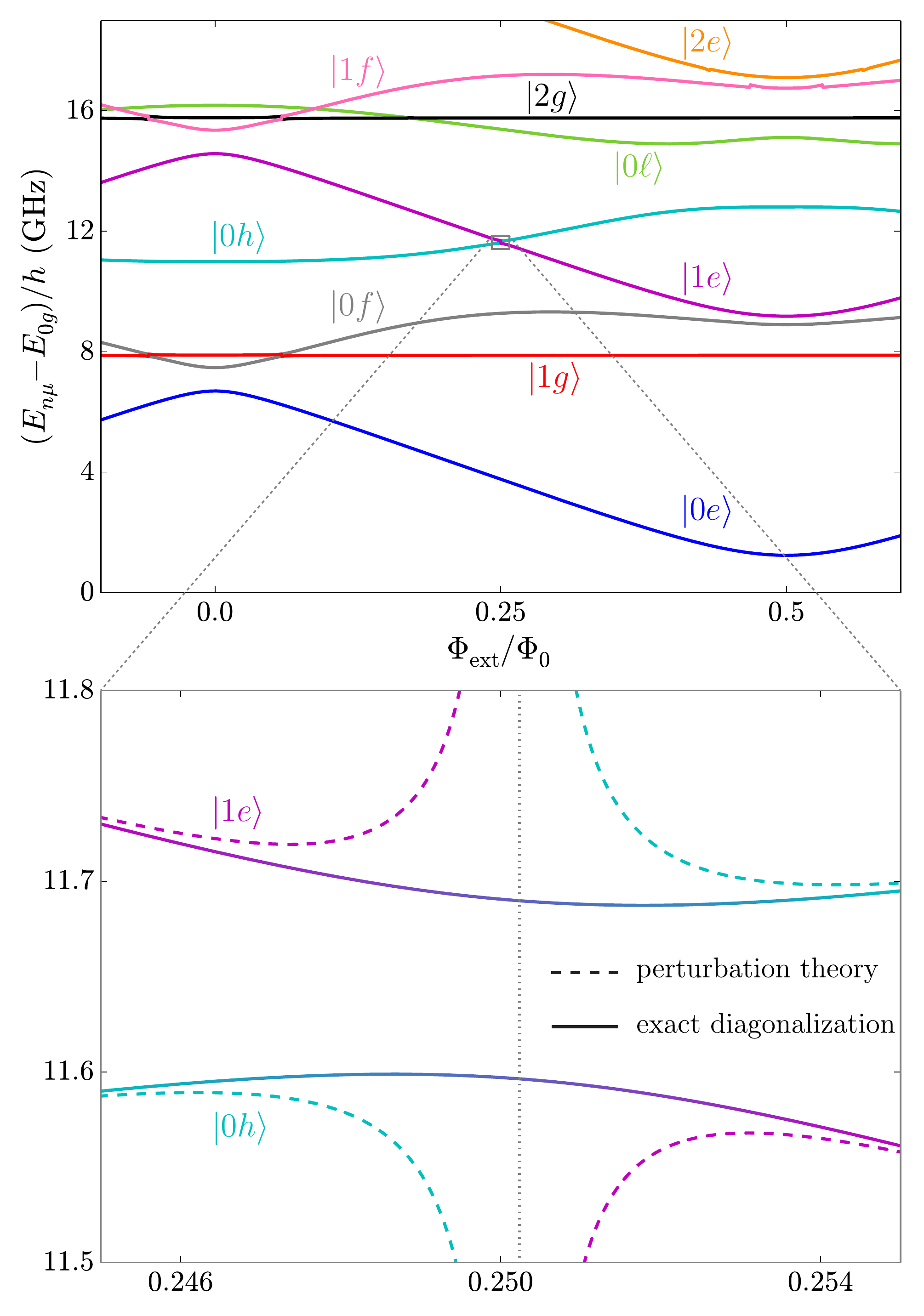}
\put(-250,340){(a)}
\put(-250,150){(b)}
\caption{(a) Energy levels $E_{n\mu}$ of the fluxonium-resonator system as a function of external flux $\Phi_\text{ext}$. The zero point energy is taken to be $E_{0g}$ and energies are given in frequency units, while flux is given in multiples of the magnetic flux quantum $\Phi_0$. (b) Anticrossing between the $0\rightarrow1$ readout transition and the $e\rightarrow h$ qubit transition, and the divergence of second-order perturbation theory.\label{fig:spectrum}}
\end{figure}

We treat the full analytic expression (\ref{eq:matrixhamiltonian}--\ref{eq:sine}) for the Hamiltonian matrix of the fluxonium-resonator system as in Fig.~\ref{fig:circuit}c (when $L_q \gg L_r \sim L_s$). Computing the lowest energy levels requires truncating both readout and qubit Hilbert spaces. This is done by restricting the basis to $\left\{ |nm\rangle : n \leq n_0, m \leq m_0\right\}$ with finite cutoff dimensions $n_0$ and $m_0$. 
We choose $n_0 \sim 5$ and $m_0 \sim 20$ in order to simultaneously minimize truncation errors and computational cost.\footnote{The computational complexity for the perturbative calculation in App.~\ref{app:perturbation} is $O(m_0^4)$ while that for exact diagonalization in Sec.~\ref{subsec:diag} is $O(n_0^3 m_0^3)$. For our system, these are on the same order.} Such numerical diagonalization yields the full solution to the time-independent Schr\"{o}dinger equation

\begin{equation} \label{eq:tise1}
\hat{H} |n\mu\rangle = E_{n\mu} |n\mu\rangle,
\end{equation}

\noindent
where $\mu$ denotes the qubit excitation. The computed energy spectrum for device A (see Tab.~\ref{tab:params}) as a function of threaded external magnetic flux $\Phi_\text{ext}$ is plotted in Fig.~\ref{fig:spectrum}a for the first nine transitions of the system from its ground state.

In order to assign the quantum numbers $n$ and $\mu$ to these energy levels, which undergo anticrossings as external flux is varied as in Fig.~\ref{fig:spectrum}, we additionally diagonalize a decoupled version of the Hamiltonian (\ref{eq:circuithamiltonian}), $\hat{H}_0$. This Hamiltonian is obtained by setting $\hat{\varphi}_R \rightarrow 0$ in the argument of the cosine in (\ref{eq:circuithamiltonian}) or by setting $c_{nn^\prime}^R \rightarrow 1$ and $s_{nn^\prime}^R \rightarrow 0$ in (\ref{eq:matrixhamiltonian}). Computing the energy levels for $\hat{H}_0$ only requires truncating the qubit Hilbert space. This time, the basis is restricted to $\left\{ |nm\rangle : m \leq m_0\right\}$ with finite $m_0$. As before, we take $m_0 \sim 20$, and we reiterate that numerical methods are the only tractable option. This procedure results in the solution of

\begin{equation} \label{eq:tise0}
\hat{H}_0 |n\mu\rangle_0 = \epsilon_{n\mu} |n\mu\rangle_0.
\end{equation}

\noindent
Note that the eigenstates are easily separable; that is, $|n\mu\rangle_0 = |n\rangle |\mu\rangle_0$ and $\epsilon_{n\mu} = \hbar \omega_R n + \epsilon_\mu$. In other words, the decoupled spectrum has built-in quantum numbers. Quantum numbers are then assigned by comparing the coupled energy spectrum $E_{n\mu}$ to the decoupled spectrum $\epsilon_{n\mu}$. In the limit of weak coupling, we may simply take $n$ and $\mu$ for a given coupled level to be those of the nearest decoupled level. In this scheme, the quantum numbers labeling a chosen energy level abruptly switch at level anticrossings, as shown in Fig.~\ref{fig:spectrum}b.


\subsection{Comparison with other methods\label{subsec:lit}}

To benchmark our method, we contrast the essential steps with those of Ref.~\onlinecite{Manucharyan2009}, \onlinecite{Nigg2012}, and \onlinecite{Zhu2013.1}. Respectively, the central differences are transforming to normal coordinates, truncating the cosine term, and using perturbation theory for the qubit-environment coupling.

\subsubsection{Normal modes}

Numerical diagonalization of the Hamiltonian corresponding to (\ref{eq:lagrangian}) in the bare harmonic oscillator basis $\{\Phi_r, \Phi_q\}$ was carried out in the case of capacitive coupling in Ref.~\onlinecite{Manucharyan2009}. Our approach performs this diagonalization in the dressed harmonic oscillator basis $\{\Phi_R, \Phi_Q\}$. This transformation allows for more aggressive Hilbert space truncation when numerically diagonalizing, provided the residual coupling mediated by the cosine is smaller than the initial linear coupling:

\begin{equation}
\frac{L_s}{L_r + L_s} < \frac{|\lambda_3|}{\lambda_4}. \label{eq:criterion}
\end{equation}

\noindent
The left-hand side in the above inequality represents the turns ratio between the shared inductance the readout inductance, while the right-hand side represents the degree of mode hybridization. Criterion (\ref{eq:criterion}) is satisfied for a broad class of systems, including our physical devices (see Tab.~\ref{tab:params}--\ref{tab:lambda}).

\subsubsection{Cosine truncation}

In Ref.~\onlinecite{Nigg2012}, the cosine term in (\ref{eq:circuithamiltonian}) is Taylor expanded and the first few terms are retained. In our treatment, no truncation is employed. This truncation makes it extremely easy to calculate the off-diagonal Hamiltonian matrix elements, facilitating numerical diagonalization or perturbative analysis. For either mode, this step requires that (i) a well-defined classical steady state $\langle \varphi_i\rangle$ is found and (ii) the quantum fluctuations of the phase $\varphi_i^\text{ZPF}$ are small. Condition (i) guarantees that the Taylor expansion is possible, and we note that it is patently violated for the fluxonium qubit, since there are multiple potential minima. Condition (ii) ensures that the expansion converges rapidly, and it mathematically amounts to

\begin{equation}
Z_i = \sqrt{\frac{L_i}{C_i}} \ll R_Q = \frac{\hbar}{(2e)^2}, \label{eq:condition}
\end{equation}

\noindent
where $Z_i$ is the characteristic impedance of mode $i$ and $R_Q$ is the impedance quantum. This highlights the fact that truncating the cosine expansion is permissible only for low impedance modes. 

\begin{table}[b!]
\caption{Parameters used for energy level calculations in Fig.~\ref{fig:spectrum}--\ref{fig:compare}.\label{tab:params}}
\begin{ruledtabular}
\begin{tabular}{c c c c}
& Device A & Device B \\
\hline
$C_r$ & $20.3 \text{ fF}$ & $20.1 \text{ fF}$ \\
$L_r$ & $15.6 \text{ nH}$ & $19.7 \text{ nH}$ \\
$C_q$ & $5.3 \text{ fF}$ & $5.9 \text{ fF}$ \\
$L_q$ & $386 \text{ nH}$ & $430 \text{ nH}$ \\
$E_J$ & $6.20 \text{ GHz}$ & $9.08 \text{ GHz}$ \\
$L_s$ & $4.5 \text{ nH}$ & $2.9 \text{ nH}$ \\
\end{tabular}
\end{ruledtabular}

\caption{Normal mode coefficients corresponding to parameter sets in Tab.~\ref{tab:params}.\label{tab:lambda}}
\begin{ruledtabular}
\begin{tabular}{c c c c}
& Device A & Device B \\
\hline
$1 - \lambda_1$ & $1.5\times10^{-3}$ & $4.1\times10^{-4}$ \\
$\lambda_2$ & $1.5\times10^{-2}$ & $8.4\times10^{-3}$ \\
$\lambda_3$ & $-5.6\times10^{-2}$ & $-2.9\times10^{-2}$ \\
$1 - \lambda_4$ & $1.1\times10^{-4}$ & $3.5\times10^{-5}$ \\
\end{tabular}
\end{ruledtabular}
\end{table}

\subsubsection{Perturbation theory}

Independently of the considerations in the previous two sections, the qubit-readout coupling terms may be treated perturbatively, as in Ref.~\onlinecite{Zhu2013.1}. This treatment greatly reduces the Hilbert space size for numerical diagonalization, at the expense of additional calculations of perturbed eigenstates and energy levels. For a variety of systems, perturbative expansions do not converge quickly enough to provide an advantage in computational complexity over numerical diagonalization (see Fig.~\ref{fig:spectrum}b as well as App.~\ref{app:perturbation} for an application to our system). Simply stated, perturbation theory is not efficient for systems with sufficiently large qubit-environment coupling.

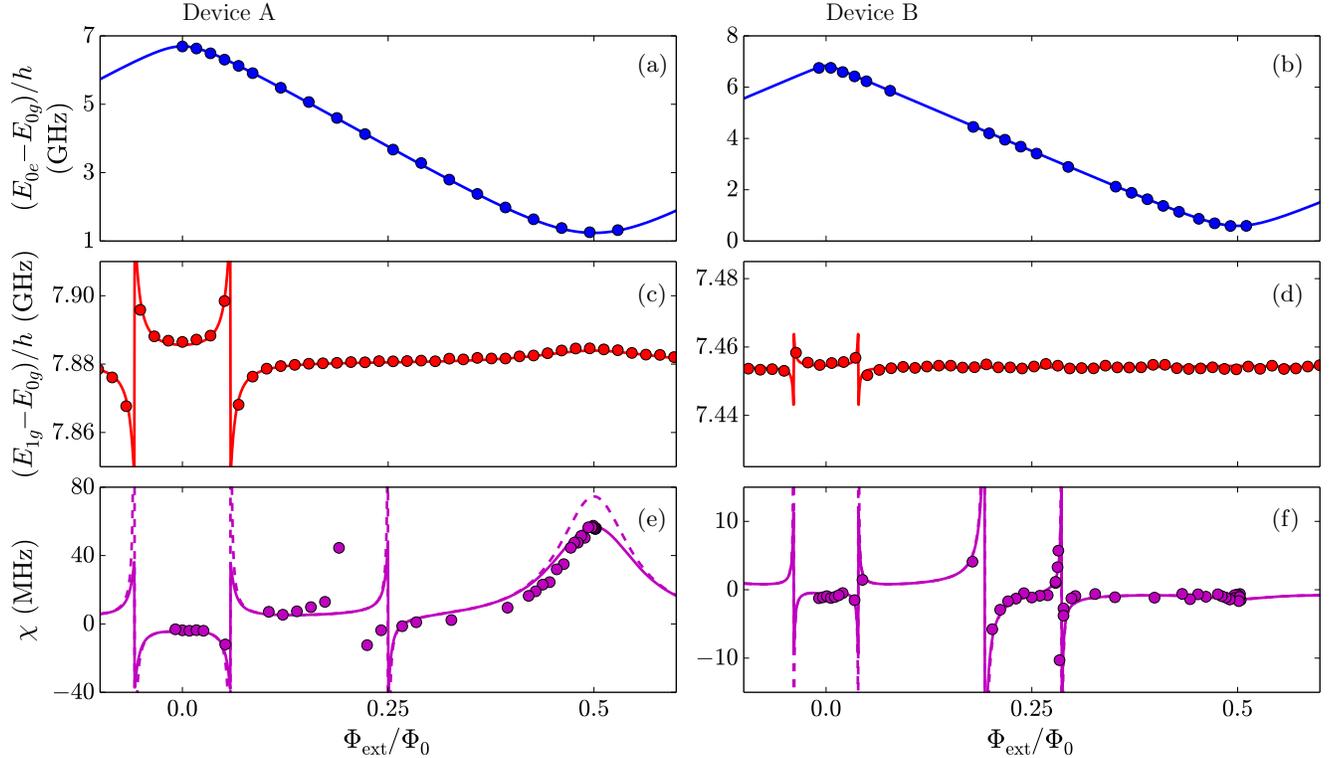
\begin{figure*}
\includegraphics[width=\textwidth]{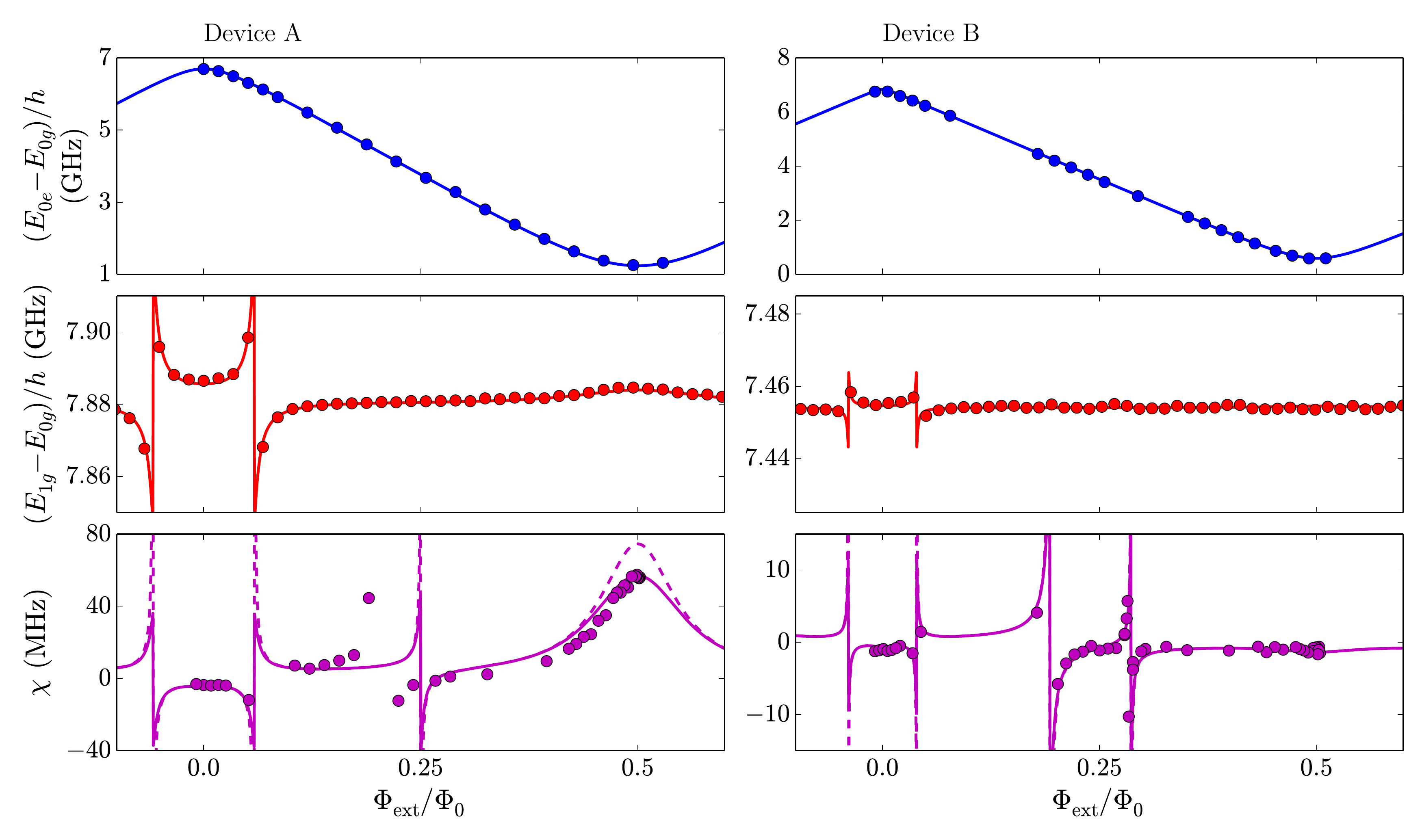}
\put(-265,267){(a)}
\put(-25,267){(b)}
\put(-265,180){(c)}
\put(-25,180){(d)}
\put(-265,95){(e)}
\put(-25,95){(f)}
\caption{(a--b) (blue): qubit $g\rightarrow e$ transition frequency. (c--d) (red): readout $0\rightarrow 1$ transition frequency. (e--f) (purple): dispersive shift $\chi$. All results are plotted as a function of external flux $\Phi_\text{ext}$ in units of $\Phi_0$. Circles indicate data taken using two-tone spectroscopy (blue), single-tone spectroscopy (red), and single-tone spectroscopy preceded by qubit state preparation (purple). Solid lines indicate theoretical fits obtained from numerical diagonalization. Dashed lines indicate results from second-order perturbation theory. \label{fig:compare}}
\end{figure*}

\section{Agreement with experiment\label{sec:experiment}}

We test the accuracy of our circuit model and analysis by comparing the simulated spectrum to experimentally obtained spectroscopy data at various values of $\Phi_\text{ext}$ and for two devices with different parameters (Tab.~\ref{tab:params}). 

These devices are measured in an impedance-matched, copper rectangular waveguide mounted in a dilution refrigerator at $\sim$\,20 mK. Two-tone microwave pulses are incident on the device and then demodulated at $300\text{ K}$ using a heterodyne interferometry setup (App.~\ref{app:expdetails}).\cite{Wallraff2004} This allows for the measurement of the readout $0\rightarrow 1$ transition frequency, the qubit $g \rightarrow e$ transition frequency, and the dispersive shift $\chi$ of the readout by the qubit. Data is shown in Fig.~\ref{fig:compare}.

It is clear that the measured readout $0\rightarrow 1$ transition frequency and the qubit $g \rightarrow e$ transition frequency should be compared to the quantities $(E_{1g} - E_{0g})/h$ and $(E_{0e} - E_{0g})/h$ obtained from diagonalization, respectively, in the limit of zero temperature. The dispersive shift $\chi$ may also be computed from diagonalization via

\begin{equation} \label{eq:chi}
\chi \equiv \frac{1}{h} \left[ \left( E_{1e} - E_{0e} \right) - \left( E_{1g} - E_{0g} \right) \right].
\end{equation}

\noindent
These three numerically computed quantities are also plotted in Fig.~\ref{fig:compare}. In addition, the dispersive shift calculated from the perturbative approach in App.~\ref{app:perturbation} is also plotted in Fig.~\ref{fig:compare}e--f.

The parameters in Tab.~\ref{tab:params} are found by fitting these simulated quantities to those measured experimentally. The readout capacitance $C_r$ is predicted independently using a commercial high-frequency electromagnetic solver (Ansys HFSS), a finite-element electromagnetic modeling program. The chief discriminating factor between device A and device B is the qubit-readout coupling strength. The turns ratio between the shared inductance and the readout unshared inductance is $L_s / L_r \approx 0.29$ for device A, while $L_s / L_r \approx 0.15$ for device B. Moreover, the value of $E_J$ is roughly $50\%$ higher for device B than device A, resulting in a significantly lower $g \rightarrow e$ transition frequency at $\Phi_\text{ext} = 0.5\Phi_0$.

Theoretical and experimental results in Fig.~\ref{fig:compare} agree well, with the exception of two features. First, the location in external flux of the singularity in $\chi$ for device A differs between the model and measurements. We attribute this to the $e \rightarrow h$ qubit transition crossing the $0 \rightarrow 1$ readout transition (see Fig.~\ref{fig:spectrum}b), which involves the $|0h\rangle$ state, whose transition frequency is $\sim$\,12 GHz from the ground state. Our approximation of the Josephson junction array composing the unshared superinductance of the qubit breaks down at these frequencies due to the fundamental mode of the array occurring at $\sim$\,11 GHz.\cite{Masluk2012,Ferguson2013,Viola2015} Second, perturbation theory consistently overestimates the dispersive shift in the vicinity of avoided crossings for both device A and (to a lesser extent) device B. This is most apparent at $\Phi_\text{ext} = 0.5\Phi_0$ for device A, at which point $\chi$ is calculated to be $75 \text{ MHz}$ using perturbation theory and $57 \text{ MHz}$ using numerical diagonalization. The error in the perturbative calculation stems from the $e \rightarrow f$ qubit transition becoming nearly resonant with the $0 \rightarrow 1$ readout transition (see Fig.~\ref{fig:spectrum}a).

The strong variation of the energy spectrum with external flux is apparent in Fig.~\ref{fig:spectrum}a.\cite{Majer2007} This indicates a valuable feature of the fluxonium-resonator system: the qubit, readout, and qubit-readout coupling all behave differently as $\Phi_\text{ext}$ is swept. As an additional example, the inherited nonlinearity of the readout resonator, extracted from the computed energy spectrum, can undergo strong variation with both qubit state and $\Phi_\text{ext}$ (App.~\ref{app:kerr}). This allows access to multiple coupling regimes and cQED Hamiltonians within a single device. 

\section{Conclusion\label{sec:conc}}

We have constructed an effective circuit (Fig.~\ref{fig:circuit}c) and Hamiltonian (\ref{eq:circuithamiltonian}) for the fluxonium-resonator system: a device consisting of a readout resonator and fluxonium qubit, where all inductances are formed by arrays of Josephson junctions and the readout-qubit coupling is realized by a partially shared inductance. The low-energy spectrum of this device has been computed using a new method: numerical diagonalization in the dressed mode basis. Quantitative agreement with experimental results was obtained for nearly all values of $\Phi_\text{ext}$. This demonstrates the utility of our method and we expect it to be applicable in diagonalizing a broad class of superconducting circuit Hamiltonians, such as strongly coupled fluxonium qubits, inductively-shunted transmon qubits, and fluxonium qubits coupled to readout resonators via shared flux qubits.

\begin{acknowledgments}
We thank G.~Catelani, S.~M.~Girvin, L.~I. Glazman, M.~Hatridge, and S.~Shankar for fruitful discussions. This work was supported by the Army Research Office Grant No.~W911NF-14-1-0011. W.C.S. was supported by DoD-MURI award No.~FP057123-C. 
Facilities use was supported by the Yale Institute of Nanoscience and Quantum Engineering under National Science Foundation Grant No.~MRS1119826.
\end{acknowledgments}

\appendix
\section{Perturbation Theory\label{app:perturbation}}

To treat the qubit-readout coupling in (\ref{eq:circuithamiltonian}) perturbatively, as shown in Fig.~\ref{fig:spectrum}b and mentioned in Sec.~\ref{sec:experiment}, we Taylor expand the cosine term about $\lambda_3 = 0$, because $|\lambda_3|\ll 1$. Discarding the $O(\lambda_3^3)$ terms results in

\begin{align} \label{eq:approxhamiltonian}
\hat{H} &= \frac{1}{2 C_R} \hat{Q}_R^2 + \frac{1}{2 L_R} \hat{\Phi}_R^2 + \frac{1}{2 C_Q} \hat{Q}_Q^2 + \frac{1}{2 L_Q} \hat{\Phi}_Q^2 \nonumber \\
&\hspace{0.27cm} - E_J \Big[ \cos \left(\lambda_4 \hat{\varphi}_Q - \varphi_\text{ext}\right) - \lambda_3 \hat{\varphi}_R \sin \left(\lambda_4 \hat{\varphi}_Q - \varphi_\text{ext}\right) \nonumber \\
&\hspace{2cm} - \frac{1}{2} (\lambda_3 \hat{\varphi}_R)^2 \cos \left(\lambda_4 \hat{\varphi}_Q - \varphi_\text{ext}\right) \Big].
\end{align}

\noindent
Although the same assumption ($|\lambda_3| \ll 1$) justifies both the replacement of (\ref{eq:circuithamiltonian}) by (\ref{eq:approxhamiltonian}) and the use of perturbation theory on the last two terms in (\ref{eq:approxhamiltonian}), the majority of error arises from the latter step. We may then consider the first five terms in (\ref{eq:approxhamiltonian}) as the unperturbed Hamiltonian $\hat{H}_0$, which coincides with the decoupled Hamiltonian from Sec.~\ref{subsec:diag}. First- and second-order perturbation theory may then be used on the seventh and sixth terms in (\ref{eq:approxhamiltonian}), respectively. This results in corrections of the form

\begin{align} \label{eq:corrections}
\delta\epsilon_{n\mu} &= E_J (\lambda_3 \varphi_R^\text{ZPF})^2 \left(n + \frac{1}{2}\right) {}_0\langle \mu | \cos \left(\hat{\varphi}_Q - \varphi_\text{ext} \right) | \mu \rangle_0 \nonumber \\
&\hspace{0.5cm} + E_J^2 (\lambda_3 \varphi_R^\text{ZPF})^2 \sum\limits_{\mu^\prime \neq \mu} \frac{(2n + 1) (\epsilon_\mu - \epsilon_{\mu^\prime}) + \hbar\omega_R}{\left( \epsilon_\mu - \epsilon_{\mu^\prime}\right)^2 - \left(\hbar \omega_R\right)^2} \nonumber \\
&\hspace{2.25cm} \times \left| {}_0\langle \mu^\prime | \sin \left(\hat{\varphi}_Q - \varphi_\text{ext} \right) | \mu \rangle_0 \right|^2,
\end{align}

\noindent
completing the calculation of the energy levels $\epsilon_{n\mu} + \delta\epsilon_{n\mu}$ for the system via perturbative treatment of the readout-qubit coupling.

\section{Experimental details\label{app:expdetails}}

The two devices in Sec.~\ref{sec:experiment} are fabricated with aluminum on sapphire substrates using double-angle evaporation. In particular, bridge-free fabrication is used for the Josephson junctions.\cite{Lecocq2011} The rectangular waveguide enclosing the resulting devices has $0.3 \text{ dB}$ insertion loss across frequencies in the passband of $6.5-9 \text{ GHz}$. An applied external magnetic flux is threaded through the qubit (Fig.~\ref{fig:circuit}b--c) by passing a current through a superconducting coil around the mid-section of the waveguide. Thermally anchored to the mixing chamber of a dilution refrigerator, this sample holder is magnetically isolated from its environment using high-permeability metallic and aluminum shields.

Input microwave signals are channeled into the waveguide through $63\text{ dB}$ of attenuation as well as infrared-frequency filters.\cite{Rigetti2012,Geerlings2013} Output signals are passed through two isolators and then amplified using a high-electron-mobility transistor and a commercial microwave amplifier at $300 \text{ K}$. Single-tone spectroscopy using a vector network analyzer is used to determine the readout $0\rightarrow1$ transition frequency. We use two-tone spectroscopy to measure the qubit $g \rightarrow e$ transition frequency. This involves sending a $\sim$\,100 $\mu$s saturation pulse at a variable qubit frequency followed by a $\sim$\,30 $\mu$s readout pulse at the fixed $0\rightarrow 1$ transition frequency. Our room temperature heterodyne interferometry setup independently mixes the outgoing readout pulse as well as a reference readout pulse with local oscillator signals (at a frequency detuned from the readout by $50 \text{ MHz}$) and then digitizes and analyzes the $50 \text{ MHz}$ components. Finally, the dispersive shift $\chi$ is measured by applying a $\pi$-rotation pulse at the qubit $g\rightarrow e$ transition frequency and then performing pulsed single-tone spectroscopy of the $0\rightarrow 1$ readout transition with the heterodyne interferometry setup. This is compared to the result from the same procedure with an off-resonant $\pi$-pulse to determine $\chi$.

\section{Inherited anharmonicity of the readout\label{app:kerr}}

\begin{figure}
\includegraphics[width=0.48\textwidth]{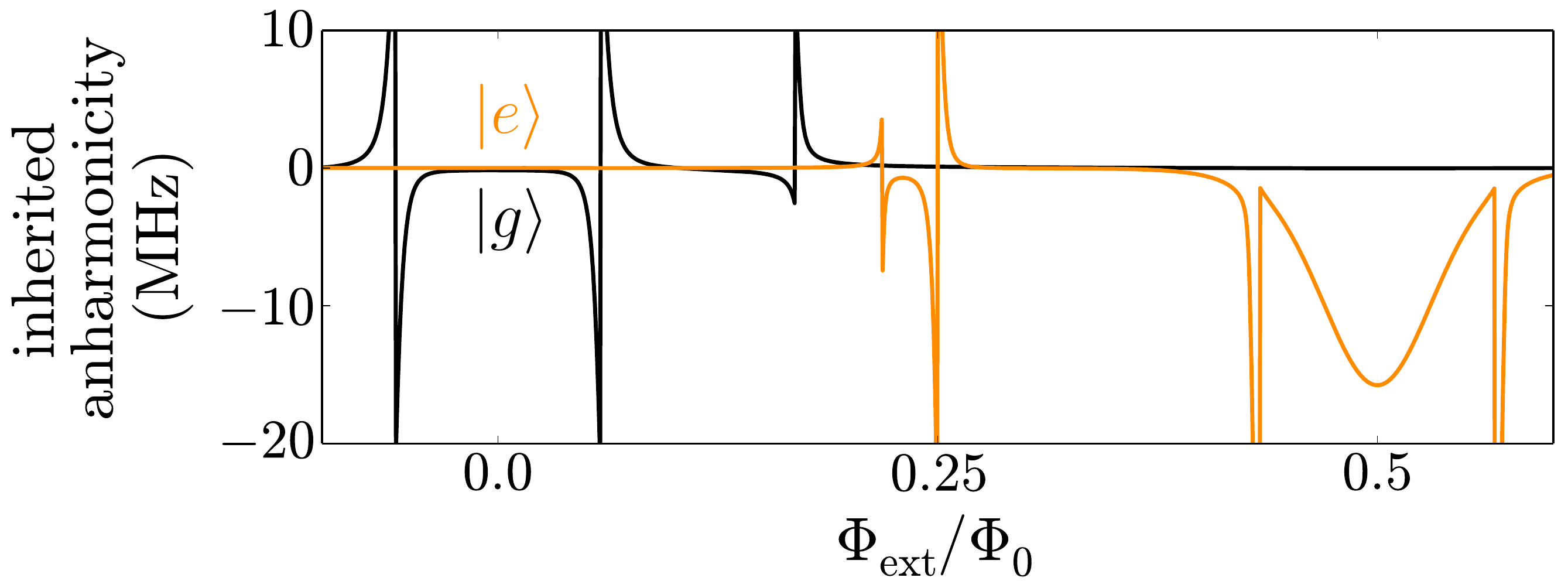}
\caption{Inherited anharmonicity of the readout resonator as a function of external flux $\Phi_\text{ext}$, defined as the difference between the $1\rightarrow 2$ and $0 \rightarrow 1$ readout transition frequencies. Results from numerical diagonalization are shown for the qubit in state $|g\rangle$ (black) and state $|e\rangle$ (orange). \label{fig:kerr}}
\end{figure}

Further proof of the tunability of the fluxonium-resonator system as a function of $\Phi_\text{ext}$ is demonstrated in Fig.~\ref{fig:kerr}. The calculated qubit state-dependent inherited anharmonicity of the readout is shown as a function of external flux for device A. We define this inherited anharmonicity as

\begin{equation} \label{eq:inheritedanharmonicity}
\frac{1}{h} \left[ \left(E_{2\mu} - E_{1\mu}\right) - \left(E_{1\mu} - E_{0\mu}\right)\right],
\end{equation}

\noindent
that is, analogously to the self-Kerr effect.\cite{Haroche2006,Kirchmair2013} Remarkably, this inherited anharmonicity changes in sign and order of magnitude for both fixed qubit state and variable $\Phi_\text{ext}$, and variable qubit state and fixed $\Phi_\text{ext}$.

\section{Input-output theory\label{app:scattering}}

In Sec.~\ref{sec:theory}--\ref{sec:experiment}, we have treated the fluxonium-resonator system as a closed quantum system; experimentally, however, the system is allowed to exchange photons with its waveguide environment. In this appendix, we will attempt to understand quantitatively how the system responds to microwaves traveling through the waveguide.

\begin{figure}
\includegraphics[width=0.5\textwidth]{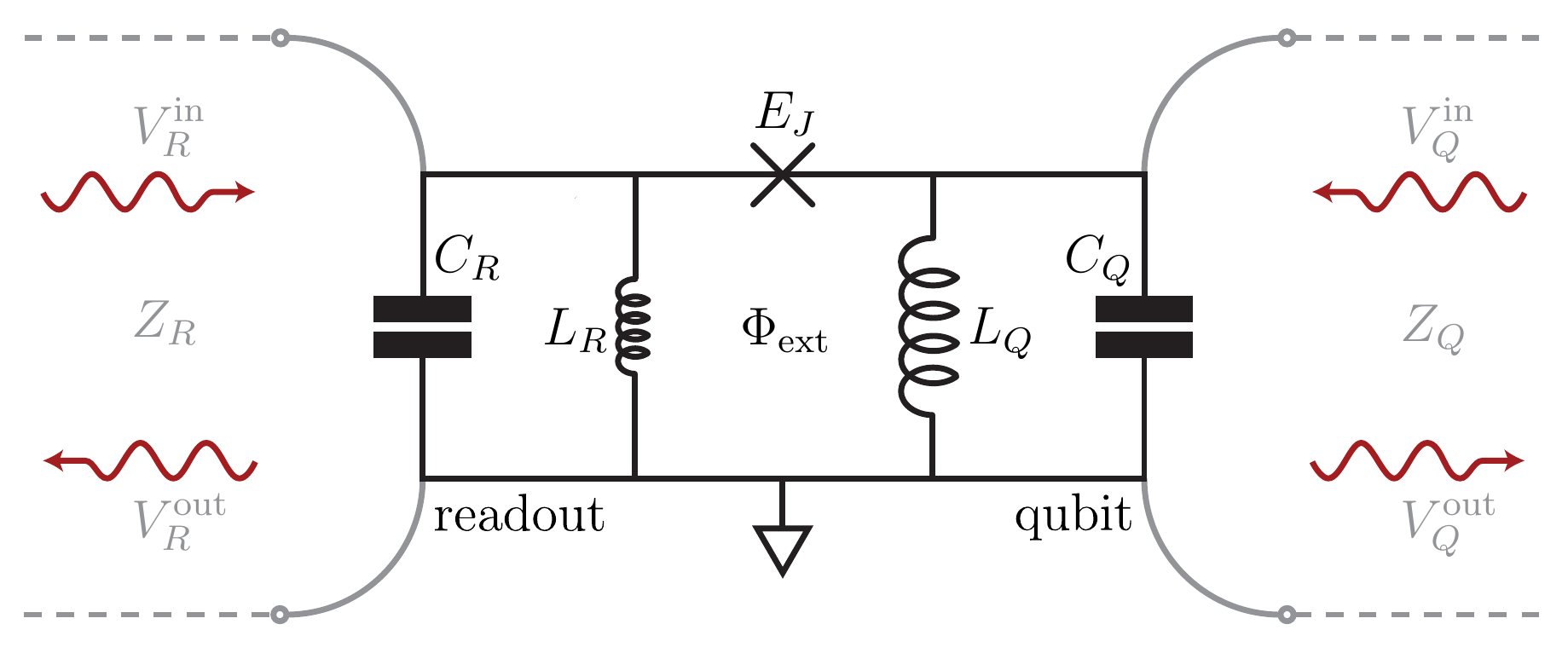}
\caption{Electrical circuit schematic for the fluxonium-resonator system (black), showing the decomposition into normal modes coupled by a single nonlinear inductive element. Transmission lines (dashed grey) are weakly coupled (solid grey) to these modes in order to treat the system as a two-port microwave device. \label{fig:portedcircuit}}
\end{figure}

Depending on its frequency and the relative electric dipole moments of the readout and qubit normal modes, a microwave pulse incident on the fluxonium-resonator system may excite either the readout or qubit mode. We therefore couple transmission lines with impedances $Z_R$ and $Z_Q$ to the readout and qubit modes, respectively, as shown in Fig.~\ref{fig:portedcircuit}.\footnote{Instead of coupling transmission lines to the circuit's normal modes, we might have chosen the transmission lines to coincide with the physical input and output ports of the waveguide. These two schemes are related by a linear transformation corresponding to the electric dipole moments of the modes, and therefore convey identical information.} Note that the electrical circuit for the system (solid black in Fig.~\ref{fig:portedcircuit}) is drawn according to the Hamiltonian (\ref{eq:circuithamiltonian}) with scaled coordinates,\footnote{Mathematically, this amounts to transforming $\lambda_3 \hat{\varphi}_R \longrightarrow - \hat{\varphi}_R$ and $\lambda_4 \hat{\varphi}_Q \longrightarrow \hat{\varphi}_Q$, in order to remove the $\lambda$ dependence in the argument of the cosine.} so that the inter-mode coupling is entirely captured by a single nonlinear inductive element. This construction, which considers both transmission lines as the waveguide, allows us to treat dissipation from the system using the two-port microwave scattering matrix approach as opposed to the Lindblad master equation formalism.\cite{Pozar2005} 

We proceed by defining the $2 \times 2$ impedance matrix $Z$ as that whose elements $Z_{ij}$ are given by

\begin{equation} \label{eq:ohm}
V_j = I_i Z_{ij},
\end{equation}

\noindent
in accordance with Ohm's law, where $V_j$ is the voltage measured across mode $j$ (with $R$ and $Q$ corresponding to readout and qubit, respectively, as in Fig.~\ref{fig:portedcircuit}) and $I_i$ is the current imposed across mode $i$. In other words, $Z_{ij}$ is the impedance response of the voltage across mode $j$ to a flux across mode $i$. Taking a time derivative in the Fourier domain, we have

\begin{figure*}
\includegraphics[width=0.9\textwidth]{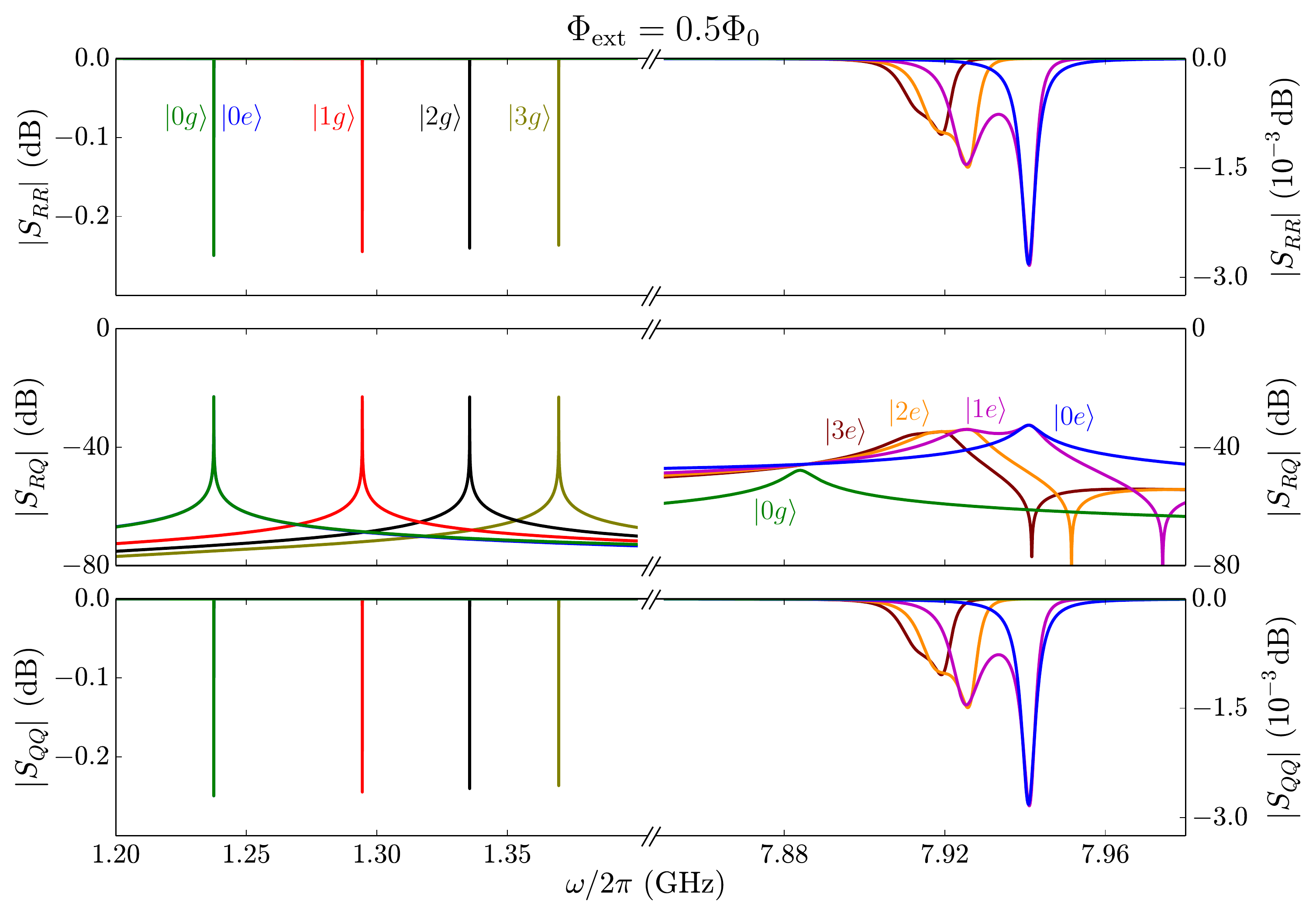} 
\put(-470,280){(a)}
\put(-470,186){(b)}
\put(-470,92){(c)}
\caption{State-dependent scattering parameter magnitudes for the fluxonium-resonator system at $\Phi_\text{ext} = 0.5 \Phi_0$. (a) Reflection coefficient amplitude for the readout mode. (b) Transmission coefficient amplitude. (c) Reflection coefficient amplitude for the qubit mode. Amplitudes are plotted at frequencies near the $g \rightarrow e$ qubit transition (left-hand side) as well as those near the $0 \rightarrow 1$ readout transition (right-hand side). Plots are layered from top to bottom in order of ascending energy. \label{fig:scattmag}}
\end{figure*}

\begin{figure*}
\includegraphics[width=0.935\textwidth]{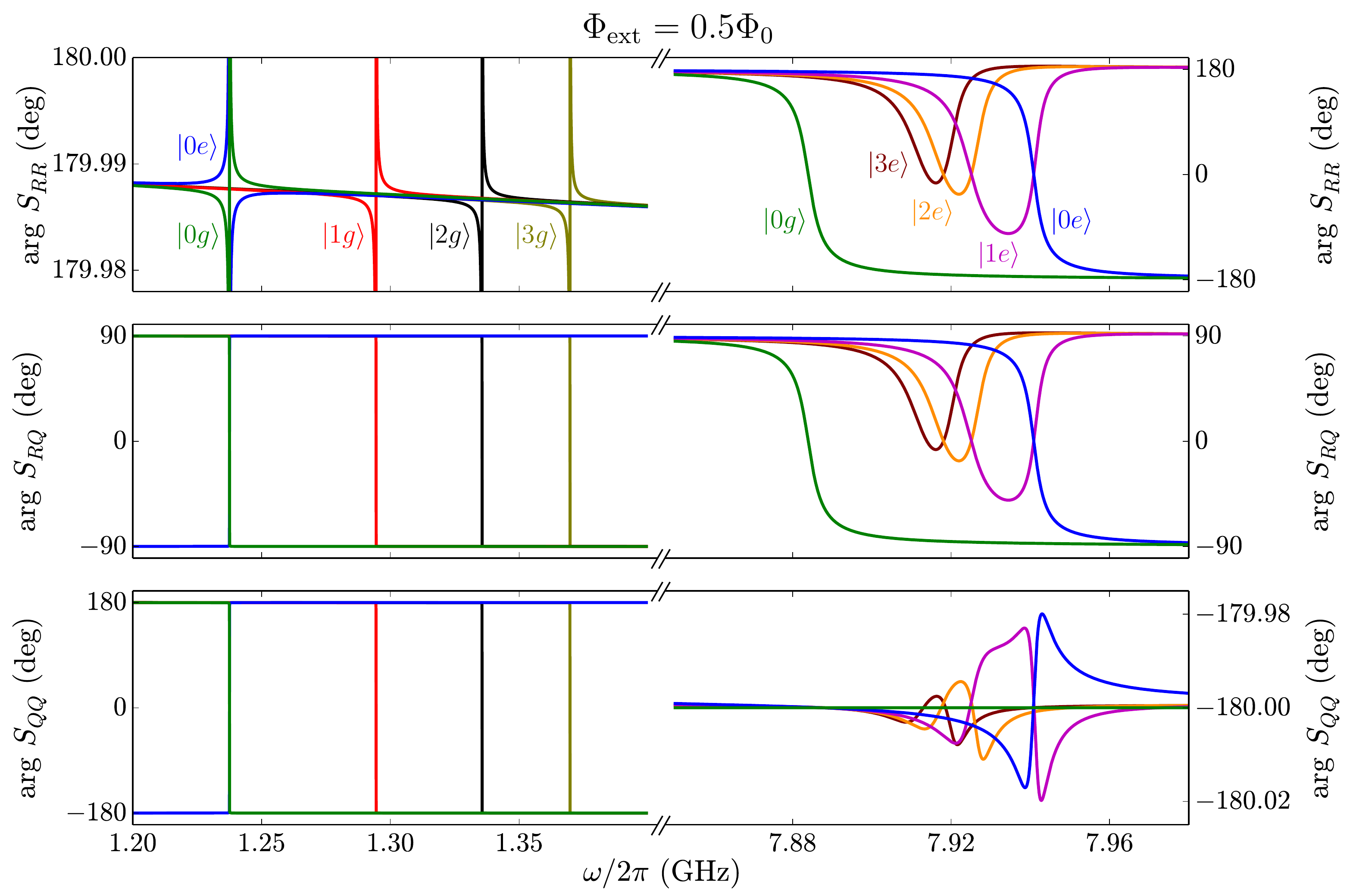}
\put(-490,288){(a)}
\put(-490,194){(b)}
\put(-490,100){(c)}
\caption{State-dependent scattering parameter arguments for the fluxonium-resonator system at $\Phi_\text{ext} = 0.5 \Phi_0$. (a) Reflection coefficient phase for the readout mode. (b) Transmission coefficient phase. (c) Reflection coefficient phase for the qubit mode. Phases are plotted at frequencies near the $g \rightarrow e$ qubit transition (left-hand side) as well as those near the $0 \rightarrow 1$ readout transition (right-hand side). Plots are layered from top to bottom in order of ascending energy. \label{fig:scattang}}
\end{figure*}

\begin{equation} \label{eq:response}
Z_{ij} = \chi_{\Phi_i V_j} = i\omega \chi_{\Phi_i \Phi_j}
\end{equation}

\noindent
in the language of linear response theory, where $\omega$ is the probe frequency. Employing the spectral expansion of the Kubo formula, we may write

\begin{equation} \label{eq:impedance}
Z_{ij} = \sum\limits_{n,\mu} \pi_{n\mu} Z_{ij}^{n\mu},
\end{equation}

\noindent
where $\pi_{n\mu}$ denotes the population of the state $|n\mu\rangle$.\cite{Kubo1957,Giuliani2005} On the other hand, $Z_{ij}^{n\mu}$ represents the impedance response of the system perfectly prepared in the state $|n\mu\rangle$, and it may be expressed as

\begin{align} \label{eq:kubo}
Z_{ij}^{n\mu} &= 2 i \omega \sum\limits_{n^\prime, \mu^\prime} \frac{E_{n^\prime \mu^\prime} - E_{n \mu}}{\left(E_{n^\prime \mu^\prime} - E_{n\mu}\right)^2 - \left(\hbar \omega\right)^2} \nonumber \\
&\hspace{2.5cm} \times \langle n\mu | \Phi_i | n^\prime \mu^\prime \rangle \langle n^\prime \mu^\prime | \Phi_j | n\mu\rangle.
\end{align}

\noindent
This expression highlights two crucial facts: the impedance response for the state $|n\mu\rangle$ has poles whenever $\omega$ resonates with the $n\mu \rightarrow n^\prime \mu^\prime$ transition and the residues of these poles scale with the matrix elements of the normal mode fluxes between the initial and final transitional states.

The $2\times 2$ scattering matrix $S$ is defined by

\begin{equation} \label{eq:scattering}
\begin{pmatrix} V_R^\text{out} \\ V_Q^\text{out} \end{pmatrix} = S \begin{pmatrix} V_R^\text{in} \\ V_Q^\text{in} \end{pmatrix},
\end{equation}

\noindent
where $V_i^\text{in}$ and $V_i^\text{out}$ depict incoming and outgoing voltage waves for mode $i$ (see Fig.~\ref{fig:portedcircuit}). We note that $V_i = V_i^\text{in} + V_i^\text{out}$ and that the characteristic impedance of the transmission line connected to mode $i$ is defined so that $V_i^{\text{in}/\text{out}} = I_i^{\text{in}/\text{out}} Z_i$, again invoking Ohm's law. Here, $I_i^\text{in}$ and $I_i^\text{out}$ denote incoming and outgoing current waves for mode $i$, where $I_i = I_i^\text{in} - I_i^\text{out}$. These relations allow us to write

\begin{equation} \label{eq:scatteringimpedance}
S = \left(Z Z_0^{-1} + I \right)^{-1} \left(Z Z_0^{-1} - I \right),
\end{equation}

\noindent
where $Z_0 = \left(\begin{smallmatrix} Z_R & 0 \\ 0 & Z_Q \end{smallmatrix}\right) $ is the characteristic impedance matrix. In our treatment, $Z$ is purely imaginary (see (\ref{eq:kubo})), and so dissipation via the coupled ports is included by choosing finite characteristic impedances. In particular, we may take $Z_i = Q_i \sqrt{L_i / C_i}$ so that the connected transmission line is matched to the impedance of its normal mode up to multiplication by $Q_i$, the quality factor of the mode. Choosing $Q_R = 1.5\times10^3$ and $Q_Q = 7.5\times10^5$, which correspond to a readout resonator linewidth of $\sim$\,5 MHz and qubit relaxation time of $\sim$\,100 $\mu$s at $\Phi_\text{ext} = 0.5\Phi_0$ for device A, we may calculate the state-dependent scattering parameters plotted in Fig.~\ref{fig:scattmag}--\ref{fig:scattang}.

Overall, we see that $|S_{ii}| \approx 1$ (near perfect reflection) and $|S_{RQ}| \approx 0$ (near vanishing transmission), which correspond to our choice of fairly large quality factors and hence weak coupling to the waveguide. On the other hand, $\arg S_{ii}$ and $\arg S_{RQ}$ experience full $360^\circ$ and $180^\circ$ rolls, respectively, at each transition frequency -- the exception being that the phase response is very small for $S_{ii}$ near its converse mode transition frequencies, which is due to the weakness of the readout-qubit coupling. 

The state-dependent scattering matrix completely characterizes the fluxonium-resonator system in the idealized waveguide environment (as modeled by transmission lines connected to constituent modes). In Fig.~\ref{fig:scattmag}--\ref{fig:scattang}, we clearly see readout resonator photon number splitting of the $g \rightarrow e$ qubit transition, a frequency shift of $\chi$ in the $|0g\rangle$ and $|0e\rangle$ phase rolls, and a shift in the readout transition frequency depending on the photon number (a manifestation of the self-Kerr effect). We note that more complicated loss mechanisms, such as internal losses, can be added to our model by adding a real part to the impedance matrix. Qualitatively, this will draw the reflection coefficient magnitudes away from unity toward zero (in the limit of critical coupling).

We note that the experimental setup used for the measurements in Sec.~\ref{sec:experiment} does not allow for complete access to the photons leaking out of the qubit mode. However, in fluorescence experiments that intentionally couple the microwave environment to the qubit, we expect these state-dependent scattering parameters to agree well with measurements.

\bibliography{paper_bib}

\end{document}